\documentstyle[aps,prl,twocolumn]{revtex}
\def\be{\begin{equation}}	
\def\ee{\end{equation}}
\def\bea{\begin{eqnarray}}
\def\eea{\end{eqnarray}}
\def\bma{\begin{mathletters}}
\def\ema{\end{mathletters}}
\def\0{\tilde{0}}
\def\1{\tilde{1}}
\def\C{{\cal C}}

\def\D{{\cal D}}
\def\A{{\cal A}}
\def\B{{\cal B}}
\newcommand{\bra}[1]{\mbox{$\langle #1 |$}}
\newcommand{\ket}[1]{\mbox{$| #1 \rangle$}}
\newcommand{\braket}[2]{\mbox{$\langle #1  | #2 \rangle$}}
\newcommand{\proj}[1]{\ket{#1}\!\bra{#1}}
\newcommand{\tr}{\mbox{tr}}

\begin{document}

\draft

\title{Optimal estimation of quantum dynamics}
\author{A. Ac\'\i n$^1$, E. Jan\'e$^{1,3}$, G. Vidal$^2$}
\address{
$^1$ Departament d'Estructura i Constituents de la Mat\`eria, Universitat de Barcelona, E-08028 Barcelona, Spain.\\
$^2$ Institut f\"ur Theoretische Physik, Universit\"at Innsbruck, A-6020 Innsbruck, Austria.\\
$^3$ Optics Section, The Blackett Laboratory, Imperial College, London SW7 2BW, UK.}
\date{\today}

\maketitle

\begin{abstract}
We construct the optimal strategy for the estimation of an unknown unitary transformation $U\in SU(d)$. This includes, in addition to a convenient measurement on a probe system, finding which is the best initial state on which $U$ is to act. When $U\in SU(2)$, such an optimal strategy can be applied to simultaneously estimate both the direction and the strength of a magnetic field, and shows how to use a spin 1/2 particle to transmit information about a whole coordinate system instead of only a direction in space.

\end{abstract}

\pacs{PACS Nos. 03.67.-a, 03.65.Bz}

\bigskip
 Consider an experimental device $\D$ that implements an unknown unitary operation $U\in SU(d)$. A probe subsystem $\A$, which can be entangled with a second subsystem $\B$, is introduced in $\D$ and analyzed at its releasing. Suppose that arbitrary manipulation is allowed on the global composite system both at the preparation and analysis stages, while $\D$ is regarded as a black box. This paper addresses the question: ``Which is the best way of estimating the operation $U$?''

 The optimal estimation of the state of a quantum system has received a lot of attention in recent years \cite{state1,state2,state3}. A situation repeatedly considered in the literature is that of a spin $1/2$ system prepared in an unknown pure state $\ket{\psi} \in \C^2$. By means of an optimal measurement on the system, the maximal amount of information about $\ket{\psi}$ is retrieved. Here we focus, instead, on the estimation of the dynamics of a quantum system (see also \cite{preskill}). This is done by analyzing, again through an adequate measurement, the changes that the initial state $\ket{\psi_0}\in\C^d\otimes\C^d$ of the system undergoes under the unknown evolution, $U\in SU(d)$. But contrary to what happens in state estimation, where only optimal measurements need to be constructed, the optimal estimation of transformations requires a double maximization: first, and most novel, we need to find the state $\ket{\psi_0}$ of the composite system that best captures the information of the transformation (unitary evolution $U$); and second, a measuring strategy that optimally retrieves such information from $U\otimes I_B\ket{\psi_0}$, where $I$ stands for the identity operator. 

 Not surprisingly, the optimal estimation of quantum transformations---necessarily based on the possibility of encoding them on, and analyzing them from, a quantum system---is closely related to the capacity of quantum systems to carry information. Our results also give insight into the role entanglement plays at enhancing the capabilities of a quantum channel: it turns out that unitary transformations are optimally encoded in the quantum correlations between the two subsystems, $\A$ and $\B$, and that, for instance, information about a whole coordinate system $\{\hat{e}_x,\hat{e}_y,\hat{e}_z\}$ can be transmited by sending only one spin $1/2$ system, provided that an ebit of entanglement between the sender and the receiver is also available. The simultaneous determination of both the direction and the strength of a magnetic field, the tuning of a quantum channel and the limits to espionage in a two-party protocol are other issues that can be addressed with the optimal scheme for the estimation of unitary operations, as we shall discuss. 

 It is easy to come up with strategies that determine $U$ with an arbitrary accuracy provided that the black-box device $\D$ can be used without restrictions. Here we are interested in the opposite situation, namely when $\D$ is used to perform the transformation $U$ only a reduced number of times $N$. We will first present an exhaustive analysis, comprising the optimal initial state $\ket{\psi_0}$ and the optimal measurement, for the case when $\D$ can only be used once, $N=1$. For the general $N$ case, and assuming that $\D$ performs the transformations in the form $U^{\otimes N}$, we will derive the optimal initial state of the system, and report the optimal POVM for $N=2$, $U\in SU(2)$.

 We start by shortly reviewing some of the elements involved in quantum estimation strategies. First, a prior probability distribution $f(U)$ uniform with respect to the Haar measure \cite{haar} expresses the fact that nothing is known about $U$ before resorting to $\D$, except that it corresponds to a unitary evolution. Second, once the device $\D$ has performed $U$ on the probe $\A$, a positive operator-valued measurement (POVM) on $\A$ and the (possibly) entangled system $\B$ will extract the information about $U$. Such POVM is a set $\{G_r\}$ of positive operators satisfying $\sum_r G_r=I_{AB}$. And third, we need a notion of how efficient a particular strategy---that is, an initial probe state $\ket{\psi_0}$ and a POVM $\{G_r\}$---is, so that we can search for the best one. There are several ways of evaluating the strategies, and the optimal solution may depend on the particular election we make. One of the main results of this paper is to present the optimal probe state $\ket{\psi_0}$ and to show that it is the same for a large class of figures of merits. Nevertheless, in order to optimize the POVM we will consider a specific, fidelity-guided figure of merits, in which the outcome $r$ of the POVM, corresponding to the operator $G_r$, is followed by a guess $U_r$ for the unknown $U$. We have chosen the function
\be
F(U, U_r) \equiv | \int_{\psi}\bra{\phi} U_r^{\dagger}U\ket{\phi}|^2 = \frac{1}{d^2} |{\rm tr} (UU_r^{\dagger})|^2 
\label{fid}
\ee
to evaluate the guess $U_r$. It quantifies, on average over all states $\ket{\phi}$, how well $U_r$ compares to $U$ when transforming $\ket{\phi}$, since it averages the overlap between $U_r\ket{\phi}$ and $U\ket{\phi}$
%\cite{fidelitat}
. Bellow we will give another interpretation to this fidelity, whose average over outcomes and unknown operations reads
\be
\bar{F} \equiv \sum_r \int_{SU(d)}f(U)dU P_r(U)F(U,U_r),
\label{avefid}
\ee
where $P_r(U)$ is the probability that the POVM produces the outcome $r$ when the device $\D$ has implemented the operation $U$.

 Let us suppose, then, that $\D$ is to be used only once. Lemma 1 presents the optimal initial state of the probe for this case. It only assumes a covariantly averaged figure of merits as in (\ref{avefid}), but where $F(U,U_r)$ is {\em any} function $h(UU_r^{\dagger})$ depending on $U$ and $U_r$ through $UU_r^{\dagger}$. Notice that only pure states need to be considered for the probe system, due to the linearity of $P_r(U)$ in the initial state (see eq. (\ref{aveh})). Therefore we take, without loss of generality, a composite probe $\A\B$, where $\A$ is the $d-$level system on which $U$ will be performed and $\B$ is a second $d-$level system, possibly entangled with $\A$.

{\bf Lemma 1:} The optimal initial state for estimating $U$ after a single performance can be chosen to be a maximally entangled state, such as 
\be
\ket{\Phi} \equiv \frac{1}{\sqrt{d}}\sum_{i=1}^d \ket{i_Ai_B}.
\label{initial1}
\ee

The reason is that, as we next show, the state $U\otimes I_B\ket{\Phi}$  can be transformed, independently of $U$, into any other state $U\otimes I_B\ket{\psi_0}$---actually, to an equally efficient state, see below---by just manipulating system $\B$.

{\bf Proof}: Let us consider the Schmidt decomposition of the most general initial state $\ket{\psi_0} \equiv \sum_{i=1}^d \lambda_i \ket{\mu_{i}\nu_{i}}$, $\lambda_i\geq \lambda_{i+1}\geq 0$, $\sum_i \lambda_i^2=1$.
We first show that the Schmidt basis $\{\ket{\mu_i\nu_i}\}$ is irrelevant as far as the average fidelity
\be
\bar{h}\equiv  \sum_r \tr \left( G_r \int dU U\!\otimes\!I_B\proj{\psi_0} U^{\dagger}\!\otimes\! I_B~ h(UU_r^{\dagger})\right)
\label{aveh}
\ee
is concerned (here $\tr ( G_r U\!\otimes\!I_B\proj{\psi_0} U^{\dagger}\!\otimes\! I_B)$ is the probability $P_r(U)$). This is so because for any $X$ and $Y \in SU(d)$, the state $X_A\otimes Y_B \ket{\psi_0}$ leads to the same maximal $\bar{h}$, as can be seen by noting that: $(i)$ any unitary transformation $Y$ in the local basis of $\B$ can be reabsorbed in the POVM elements $G_r$, whereas $(ii)$ if we prepare $\A\B$ in state $X\otimes I_B \ket{\psi_0}$ instead of $\ket{\psi_0}$, then the shift $U\rightarrow UX$ in the integration variables $U$ of eq. (\ref{aveh}), simultaneous to a shift $U_r\rightarrow U_rX$ for the guesses leads again to the same $\bar{h}$, as a consequence of the isotropy of $f(U)$. Therefore we can take
\be
\ket{\psi_0}=\sum_{i=1}^d \lambda_i \ket{i_A i_B} = \sqrt{\frac{d}{\tr M^2}}I_A \otimes M \ket{\Phi},
\ee
where $M$ is a diagonal operator with entries $M_{ii} \equiv \lambda_i/\lambda_1 \leq 1$. Suppose now that the initial state is $\ket{\Phi}$. Then $\D$ transforms it into $U\otimes I_B \ket{\Phi}$. Let us consider a covariant POVM \cite{state2} on $\B$ given by operators $\{M_Y\equiv (d/\tr M^2)^\frac{1}{2}MY\}$, where $Y$ runs isotropically over $SU(d)$ and $\int dY M_Y^{\dagger}M_Y = I_B$. It transforms $U\otimes I_B \ket{\Phi}$ into $(d/\tr M^2)^\frac{1}{2}U \otimes MY \ket{\Phi} = (d/\tr M^2)^\frac{1}{2}U Y^T \otimes M \ket{\Phi} = UY^T\otimes I_B\ket{\psi_0}$ for some known $Y$ [here we have used that, $\forall Y\in SU(d)$, $I_A\otimes Y\ket{\Phi} = Y^T\otimes I_B\ket{\Phi}$]. But this is as if we would have started with state $Y^T\otimes I_B \ket{\psi_0}$, which leads to the same average fidelity as $\ket{\psi_0}$. $\Box$

%Indeed, the state $U\otimes I_B \ket{\psi_0}$, resulting after $U$ has been performed on $\A$, can be further processed on $\B$ with independence of the unknown $U$. Any unitary manipulation on $\B$ can be absorbed in the POVM elements $G_r$, leaving $\bar{h}$ unchanged. But the local basis $\{\ket{\mu_i}\}$ for system $\A$ is also irrelevant, that is $W\otimes I_B\ket{\psi_0}$ leads to the same average fidelity $\bar{h}$ as $\ket{\psi_0}$ for any unitary $W$. This can be seen by simply making a shift $U\rightarrow UW$ in the integration variables $U$ of eq. (\ref{aveh}), simultaneous to a shift $U_r\rightarrow U_r^{\dagger}W$ for the guesses. Now, the state
%\be
%\int_{SU(d)} f(W)dW W\!\otimes \!I_B\ket{\psi_0}\otimes\ket{W}_C,
%\label{int}
%\ee
%---where the set $\{\ket{W}\}$, $\braket{W}{W'}=\delta(W-W')$, correspond to an infinite dimensional ancillary system $C$---can be transformed into $W\otimes I_B\ket{\psi_0}$, for some $W$, by means of a projective measurement on $C$. The same measurement leads to $UW\otimes I_B \ket{\psi_0}$ when performed after $U$ has been applied on the state (\ref{int}), and therefore the state (\ref{int}) is
% at least as good initial state as $\ket{\psi_0}$. But by applying Schur's lemma \cite{haar} to the computation of its reduced density matrix for system A,
%\be
%\rho_A = \int f(W)dW W \tr_B [\proj{\psi_0}]W^{\dagger} = \frac{I_A}{d},
%\ee
%we see that it is a maximally entangled state.$\Box$

 Let us now notice that our particular choice of fidelity, eq. (\ref{fid}), corresponds precisely to the probability $|\bra{\Phi}U_r^{\dagger}U\otimes  I_B\ket{\Phi}|^2$ that the state $U_r\otimes I_B\ket{\Phi}$ behaves as if it were $U\otimes I_B\ket{\Phi}$. Therefore $F(U,U_r)$ measures how similar the two operations $U$ and $U_r$ are by comparing two related states: those that best capture the information of both transformations after a single run of $\D$.

 Suppose finally that system $\A$, in the entangled state $\ket{\Phi}$ with system $\B$, has already been introduced in the black box $\D$, which produces the state $U\otimes I_B\ket{\Phi}$---denoted by $U\ket{\Phi}$ from now on---. Which is the best POVM that can be performed in order to learn about $U$ from this state? 
 We can rewrite the average fidelity of Eq. (\ref{avefid}) as
\be
\bar{F}_1 = \frac{1}{d^2}\sum_r \tr \left[  G_r \int\!\! f(U)dU U\proj{\Phi}U^{\dagger} |\tr (UU^{\dagger}_r)|^2 \right]. 
\label{fidN1}
\ee
By means of a shift $U \rightarrow V = U_r^{\dagger}U$ in the integration variables, each of the integrals inside the trace has the form $U_rf_1U_r^{\dagger}$, where
\bea
f_1 &\equiv& \int f(V)dV V\proj{\Phi}V^{\dagger} |\tr V|^2 \nonumber \\
&=& d^2\bra{\Phi} \int f(V)dV V^{\otimes 2}(\proj{\Phi})^{\otimes 2}V^{\dagger\otimes 2} \ket{\Phi}.
\label{f1}
\eea
Schur's lemma \cite{haar} states that this last integral is proportional to the identity in each of the two corresponding irreducible representations of $SU(d)$, namely the symmetric and the antisymmetric ones. A careful analysis [recalling that each $V$ is acting only on the first half of the corresponding $\ket{\Phi}$] and patient simple algebra leads to
\be
f_1 =\frac{1}{d^2-1} \left(\frac{d^2-2}{d^2}I_A\otimes I_B + \proj{\Phi}\right).
\label{f1r}
\ee
Thus $\ket{\Phi}$ is the eigenvector of $f_1$ with greatest eigenvalue, $\lambda_m \equiv 2/d^2$, and $\tr (~U_r^{\dagger}G_rU_r f_1~) \leq \lambda_m \tr G_r$ in eq. (\ref{fidN1}). Since $\sum_r \tr G_r = d^2$, the maximal fidelity can be $2/d^2$ at most. A covariant POVM \cite{state2} with operators and guesses given by $\{W\proj{\Phi}W^{\dagger}, W\}_{W\in SU(d)}$ reaches $\bar{F}_1=2/d^2$, which is consequently the optimal one. 
 
 This result is to be compared with the optimal fidelity $\bar{F}_0= 1/d^2$ made by blindly proposing a unitary transformation, say $I$ (or any other):
\be
\int f(U)dU  \frac{|\tr U|^2}{d^2} = \bra{\Phi} \int dU f(U) U\proj{\Phi}U^{\dagger} \ket{\Phi}
\ee
(the last integral is simply $I/d^2$ because of the Schur's lemma) and also with the {\em separable} fidelity $F_1^{sep} = (d+2)/[(d+1)d^2]$, which is the best fidelity that can be achieved without entangling $\A$ and $\B$, and can be computed using eq. (\ref{f1}) and the fact that a pure state of $\A$, say $\ket{0}$, is $\sqrt{d}\braket{0_B}{\Phi}$. Finally, we note that a finite (and thus physical) optimal measurement, actually one with the minimal number of outcomes, consists in a von Neumann measurement on a basis of $d^2$ maximally entangled states. For instance, on the Bell basis, with guesses $I,i\sigma_x,i\sigma_y$ and $i\sigma_z$, for the $SU(2)$ case \cite{preskill}. This completes the analysis of $N=1$ \cite{elsewhere}.

 Let us discuss some applications of the previous results. Consider first the group SU(2). Our optimal strategy can be readily applied to determine a constant magnetic field $\vec{B}= B\hat{m}$ by using the magnetic moment of a spin 1/2 particle, say an electron. Let $H_{int} = \vec{\mu}\cdot \vec{B}$ be the interaction Hamiltonian, where $\vec{\mu} = \mu (\sigma_x,\sigma_y,\sigma_z)$ and all physical constants have been absorbed in $\mu$. Then after a time $T$ the spin has evolved according to $\exp(-i\mu BT\hat{m}\cdot\vec{\sigma})$, and therefore we can identify the direction $\hat{m}$ of the magnetic field and its intensity $B$ (actually $\mu BT$). Our results show how to {\em optimally} extract information about $\vec{B}$ by means of an electron if this interacts {\em once} with the magnetic field (see also \cite{preskill}). 
 
 In the discussion above the information about the magnetic field $\vec{B}$ is not contained in the state of the spin alone, but in the correlations between this spin and a second one. Similarly, if two distant parties, Alice and Bob, want to use a recently established $d$-dimensional quantum channel,
\be
\sum_{i=1}^d c_i \ket{i_A} \longrightarrow \sum_{i=1}^d c_i \ket{i_B},
\ee
but Bob does not know the correspondence between states---that is, he ignores the states $\{\ket{i_B}\}$---, they can benefit from a maximally entangled state $\ket{\Phi}$ in order to tune the channel. Indeed, by Alice sending her half of $\ket{\Phi}$ down the channel, Bob can estimate the whole unknown basis $\{\ket{i_B}\}$ or, equivalently, the transformation $U=\sum_i\ket{i_B}\bra{i_A}$, with a fidelity $2/d^2$, which is $2(d+1)/(d+2)$ times greater than the fidelity he could have obtained also after a single use of the channel if no entanglement would have been available.
 In a sense, this is a general manifestation of how entanglement enhances the capacity of a quantum channel, with traditional quantum super-dense coding \cite{SDC} appearing as a particular case, namely when the channel is used to transmit classical information only. 

Let us further see this in the $SU(2)$ case, by assuming that a spin $1/2$ particle is used as a channel. Here an ebit of entanglement allows to transmit, by sending a single spin $1/2$ particle, information about a whole transformation $U(\hat{n},\omega) \in SU(2)$ or, equivalently, a rotation $R(\hat{n},\omega)\in SO(3)$. In other words, instead of using the spin of the particle to try to establish a common direction $\hat{n}$ in space (that of the one-qubit pure state $\proj{\psi} = 1/2(I+\hat{n}\cdot\vec{\sigma})$), Alice can now send information about a whole coordinate system $\{\hat{e}_x,\hat{e}_y,\hat{e}_z\}$ to Bob in order to establish a common reference frame. This works as follows. The parties share the state $\ket{\Phi} = (\ket{1_A1_B}+\ket{2_A2_B})/\sqrt{2}$, where $\{\ket{i_A}\}$ and $\{\ket{i_B}\}$ are given with respect to reference frames of Alice and Bob respectively. Each party knows his/her own reference frame, but ignores the other one. If Alice sends her half of $\ket{\Phi}$ to Bob, then Bob can estimate the rotation $R(\hat{n},\omega)$ (or corresponding unitary $U=\ket{1_B}\bra{1_A} + \ket{2_B}\bra{2_A}$) that relates the two coordinate frames.

Another scenario in which these results are relevant is that of two parties that are to collaborate in some task but do not trust each other. For instance, 
Bob needs to compute on a given input state $\ket{\psi}$ a function (unitary $U$) that Alice's computer can perform, but he ignores $U$. Alice is willing to assist Bob by computing $U\ket{\psi}$, but without letting him find out which transformation $U$ is. In this case Alice knows that Bob can estimate $U$ at most with a fidelity $2/d^2$. 

 So far we have analyzed a single run of the device $\D$. In practice, one would like to determine $U$ with arbitrary precision, and this is only possible if $\D$ is used many times. Suppose $U$ is performed twice. A most general strategy consists on sequentially introducing two probes, $\A_1$ and $\A_2$, on $\D$, but allowing for an arbitrary manipulation of the proofs in between. We do not know how to tackle the problem in its full generality. We will suppose {\em ad hoc} that the device $\D$ takes $N$ probes, $\A_1$...$\A_N$, and transforms them according to $U^{\otimes N}$. This could correspond, in the $SU(2)$ case, to letting the spin of $N$ electrons interact with the constant magnetic field $\vec{B}$ during some time interval $T$. 

The first step towards an optimal strategy for estimating $U$ is again to find an optimal initial state $\ket{\psi_0^N}$ for the $N$ $d$-level systems  $\A\equiv\A_1...\A_N$ and $N$ auxiliary $d$-level systems $\B\equiv\B_1...\B_N$, that lemma 2 presents. The $U^{\otimes N}$ representation of $SU(d)$ contains (several copies of) $q$ inequivalent irreducible representations (IRREPs), labeled by $\alpha=1,..., q$ in what follows. For each $\alpha$ there are $n_{\alpha}$ equivalent IRREPs, labeled by $\alpha\beta$, $\beta=1,...,n_{\alpha}$, each one having dimension $d_{\alpha}$. The set $\{\ket{\alpha\beta k}\}_{k=1}^{d_{\alpha}}$ denotes an orthonormal basis for the IRREP $\alpha\beta$, $P_{\alpha\beta}\equiv \sum_{k} \proj{\alpha\beta k}$ and $P_{\alpha} \equiv \sum_{\beta=1}^{n_{\alpha}} P_{\alpha\beta}$. The $\alpha\beta$ and $\alpha\beta'$ IRREPs being equivalent, there exists a unitary $\Pi^{\alpha}_{\beta\beta'}$ such that $U^{\otimes N}\ket{\alpha\beta k}=\Pi^{\alpha}_{\beta\beta'}U^{\otimes N}\ket{\alpha\beta' k}$ for any $U$ and $k$ \cite{haar}.

{\bf Lemma 2:} The optimal initial state for estimating $U^{\otimes N}$ is 
\be
\ket{\Phi^N} \equiv \sum_{\alpha=1}^q a_{\alpha} \ket{\Phi_{\alpha}^N},~~~\sum_{\alpha} a_{\alpha}^2 =1,
\label{optimN}
\ee
where the value of $a_{\alpha}\geq 0$ depends on the figure of merits under consideration and where
\be
\ket{\Phi^N_{\alpha}} \equiv \frac{1}{\sqrt{n_{\alpha} d_{\alpha}}} \sum_{\beta=1}^{n_{\alpha}}\sum_{k=1}^{d_{\alpha}} \ket{\alpha\beta k}_A\ket{\alpha\beta k}_B
\ee
is a maximally entangled state between the subspace of $\A$ that carries the $n_{\alpha}$ IRREPs $\alpha\beta$ (i.e., between the support of $P_{\alpha}$) and an equivalent subspace of $\B$. For instance, for the $N=d=2$ case, the optimal initial state is
\be
\ket{\Phi^2_a} \equiv a\frac{1}{\sqrt{3}} \sum_{k=1}^3 \ket{t_k}_A\ket{k}_B+\sqrt{1-a^2} \ket{s}_A\ket{4}_B,
\label{ini2}
\ee
where $\ket{t_k} \in \{\ket{00},(\ket{01}+\ket{10})/\sqrt{2}, \ket{11}\}$ are the triplet states and $\ket{s}\equiv(\ket{01}-\ket{10})/\sqrt{2}$ is the singlet state. 

{\bf Proof:} Being a generalization of that of lemma 1, here we will only sketch the proof. Notice that any state $\ket{\psi_0^N}$ of the probes can be writen as $\ket{\psi_0^N} = \sum_{\alpha=1}^q \ket{\psi^N_{\alpha}}$, where
\be
\ket{\psi^N_{\alpha}} \equiv \sum_{\beta=1}^{n_{\alpha}} \sum_{k=1}^{d_{\alpha}} \ket{\alpha\beta k}_A\ket{\phi_{\alpha\beta k}}_B
\ee
is the projection $P_{\alpha}\otimes I_B\ket{\psi^N_0}$ and $\ket{\phi_{\alpha\beta k}}$ are arbitrary states of $\B$. Since $U^{\otimes N}$ does not mix IRREPs, we can perform a global unitary transformation $V_{AB}$ that commutes with $U^{\otimes N}$ and such that we achieve $\braket{\phi_{\alpha\beta k}}{\phi_{\alpha'\beta' k'}}=\delta_{\alpha,\alpha'}\delta_{\beta,\beta'}c^{\alpha\beta}_{kk'}$, that is, the supports of $P_{\alpha\beta} \otimes I_B\ket{\psi_0^N}$ on $\B$ for different IRREPs $\alpha\beta$ and $\alpha'\beta'$ are orthogonal. For instance, in the $d=N=2$ case, where $\ket{\psi^2_t}=\sum_k \ket{t_k}_A\ket{\phi_k}_B$ and $\ket{\psi^2_s}= \ket{s}\ket{\phi}$, we can take, without loss of generality, $\braket{\phi}{\phi_l}=0$. We will now show that $\ket{\Phi^N}$ can be transformed into a state as efficient as $\ket{\psi_0^N}$ as far as the fidelity
\be
\bar{h}\equiv  \sum_r \!\tr \!\left[ G_r \!\int\! dU U^{\otimes N}\proj{\psi^N_0} U^{\dagger\otimes N} h(UU_r^{\dagger})\right]
\label{aveh2}
\ee
is concerned. This is made in two steps. First, the POVM in $\A$ defined in each $\alpha$ by $\{Q_i^{\alpha}\equiv \sum_{\beta}^{n_{\alpha}} (a_{\alpha\beta}/a_{\alpha}) \Pi^{\alpha}_{\beta,\beta\!+\!i}P_{\alpha,\beta\!+\!i}\}_{i=1}^{n_{\alpha}}$, where $\sum_i Q_i^{\alpha\dagger}Q_i^{\alpha} = P_{\alpha}$ and the sum $\beta\!+\!i$ is modulus $n_{\alpha}$, takes with certainty the state $U^{\otimes N}\ket{\Phi^N}$ into $U^{\otimes N}\ket{\Phi'}$, which is still maximally entangled in each IRREPS $\alpha\beta$ but with different weights $a_{\alpha\beta}/a_{\alpha}$ in each IRREP, where $\sum_{\beta} (a_{\alpha\beta})^2 = a_{\alpha}^2$. And second, a covariant POVM in $\B$ given by the set of operators $\{ Q_Y \equiv \sum_{\alpha}\sum_{\beta} a_{\alpha\beta}\sum_k\ket{\phi_{\alpha\beta k}}\bra{\alpha\beta k}Y^{\otimes N}\}$, where $\int dY Q_Y^{\dagger} Q_Y = I_B$ and $a_{\alpha\beta}\equiv(\sum_kc^{\alpha\beta}_{kk})^{-1/2}$, will produce, when applied on $U^{\otimes N}_A\ket{\Phi'}$, the state $(UY^T)^{\otimes N}_A \ket{\psi_0^N}$. This state corresponds to starting with $Y^{T\otimes N}_A \ket{\psi_0^N}$, which leads to the same $\bar{h}$ as $\ket{\psi_0^N}$ (see lemma 1). $\Box$   

For $N=d=2$, and by using the techniques developed in this paper, we have found that the optimal fidelity is $\bar{F}_2=(3+\sqrt{5})/8\approx 0.6545$, which corresponds to the initial state $\ket{ \Phi^2_a}$ of eq. (\ref{ini2}) with $a^2=(5+\sqrt{5})/10$ and to a covariant POVM and guesses given by $\{W^{\otimes 2}\proj{\Phi^2_{a'}}W^{\dagger\otimes 2}, W\}$, $a'^2=9/10$.

In conclusion, in this letter we have studied the optimal estimation of an unknown unitary operation, $U\in SU(d)$, when this transformation can be performed a reduced number of times, $N$. For any $N$ the best initial state has been essentially found for a large class of figures of merits. In the case of the fidelity defined in (\ref{fid}), its optimal value and the measurement that attains it are given for any dimension when $N=1$, and for $d=2$ when $N=2$. 

\bigskip

We thank L. Masanes and J.I. Cirac for useful comments. Financial support by the Spanish MEC (AP98 and AP99) and the European Community (ESF; project EQUIP; HPMF-CT-1999-00200) is acknowledged.

\end{document}